\documentclass[prb,twocolumn]{revtex4}
\usepackage{amsfonts,amssymb,amsmath}
\usepackage{graphicx}

\newcommand{\p}{\textrm{Prob}}
\newcommand{\s}[1]{|#1\rangle}

\begin{document}


\title{Entanglement, which-way measurements, and a quantum erasure}

\author{Christian Ferrari}
\affiliation{Liceo di Locarno, Via F. Chiesa 15, 6600 Locarno, Switzerland}

\author{Bernd Braunecker}
\affiliation{Department of Physics, University of Basel, Klingelbergstrasse 82, 4056 Basel, Switzerland}

\date{March 2, 2010. Accepted for American Journal of Physics}

\begin{abstract}
We present a didactical approach to the which-way experiment and the counterintuitive effect of the quantum
erasure for one-particle quantum interferences. The fundamental concept of
entanglement plays a central role and highlights the complementarity between
quantum interference and knowledge of which path is followed by the
particle.
\end{abstract}


\maketitle


\section{Introduction}

One-particle quantum interference is one of the most important effects that illustrates
the superposition principle and thus the major difference between
quantum and classical physics.\cite{F,R} In this paper we propose a simple model based on 
the Mach-Zehnder interferometer. Our hope is to provide a simple example of
quantum superposition and quantum interference.

We consider a modification of the gedanken experiment by Scully, Englert, and Walther,\cite{SEW}
which we reduce to probably the simplest setup that can expose the physics in
a concise way.
Reference~\onlinecite{SEW} is a highly influential paper and several previous publications
discuss and present the experiment in a didactical way.\cite{ESW,M,SLP,WD,G} The emphasis in these
publications ranges from practical realizations of a Mach-Zehnder interferometer
to a thorough discussion of the subtleties of quantum physics.

In this paper we show that the fundamental aspects of the experiment
can be captured by a minimal model that requires knowledge only of two-level systems and 
is based on the Mach-Zehnder interferometer. For maximum clarity we avoid an extended
discussion of experimental and further theoretical aspects, for which we refer to 
Refs.~\onlinecite{ESW,M,SLP,WD,G}.
We also focus on a Mach-Zehnder interferometer with only two detectors at the exit 
instead of the screen used in Young's two-slit experiment, which corresponds to a 
continuum of detectors.
The Mach-Zehnder interferometer allows us to model the step by step evolution of the 
state of a quantum particle
in the interferometer.

\begin{figure}
	\includegraphics[width=0.6\columnwidth]{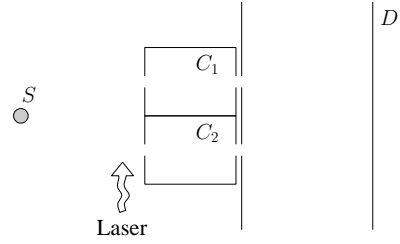}
\caption{Schematic of the experiment proposed in Ref.~\onlinecite{SEW}.
	\label{ww1}}
\end{figure}

We briefly summarize the experiment proposed in Ref.~\onlinecite{SEW} in which a 
mechanism is proposed to detect the path
(``which-way detection'') of a particle passing through a Young
interferometer (see Fig.~\ref{ww1}).
An atom is emitted by a source $S$, passes through two slits, and is detected
on a screen $D$. Directly after leaving the source, the atom is
brought into an excited state by a laser.
Two cavities $C_1$ and $C_2$ are placed in front of the slits of the Young
interferometer. When passing through the cavities the atom emits a photon and
relaxes to its ground state.
To know which path was taken by the
atom it is sufficient to see whether the photon is in $C_1$ or $C_2$.
Important for this experiment
is that the trajectory of the atom through the slits remains otherwise unperturbed.
Due to the emission of the photon in cavity $C_1$ or $C_2$ the usual
interference pattern at the screen $S$ is destroyed.
The interference disappears even without explicit
detection of the photon. It is sufficient to transfer the potential which-way
information to the photon state.
However, by allowing the photon emitted in $C_1$ or $C_2$ to be reabsorbed by an auxiliary atom,
a \emph{quantum eraser}, the information of the atom's path can be
erased, and the interference pattern at the screen can be restored.
This result was confirmed experimentally by D\"urr, Nonn, and Rempe\cite{DNR}
using a modified Mach-Zehnder interferometer (see also Refs.~\onlinecite{RY} and \onlinecite{Gogo}).

\section{The Mach-Zehnder interferometer}

\begin{figure}[h]
	\includegraphics[width=0.8\columnwidth]{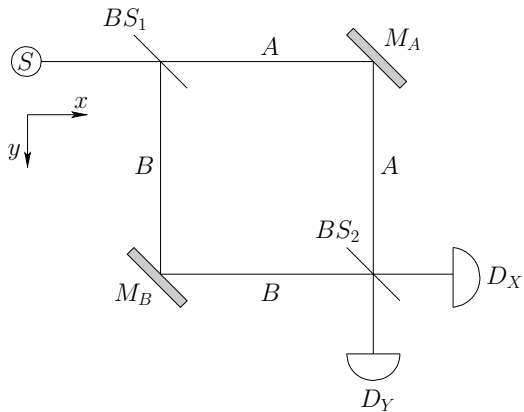}
\caption{The Mach-Zehnder interferometer.
A particle (atom) is emitted by the source $S$
and travels to the detectors $D_X$ and $D_Y$. At the beam splitter
\textit{BS}$_1$ it is deflected to path $A$ or path $B$. With the mirrors $M_A$
and $M_B$ the particle interferes with itself at the beam splitter \textit{BS}$_2$
before entering the detectors. $A$ and $B$ have the same
length, and therefore quantum interference leads to detection probability equal to 1 in $D_X$
and 0 in $D_Y$, in contrast to the classical expectation of equal probabilities of
$1/2$ for both detectors.\label{MZ1}}
\end{figure}

\noindent We consider the Mach-Zehnder interferometer shown in Fig.~\ref{MZ1}.
It consists of a source $S$, which emits particles into the
interferometer along the $x$ direction such that at any given time, a maximum of a single particle is
in the interferometer. (See Ref.~\onlinecite{R} for a discussion of the first experiment realizing
single-particle interference.) The particles first hit beam splitter \textit{BS}$_1$ through which
they are transmitted to path $A$ or deflected to path $B$. Two mirrors
$M_A$ and $M_B$ let these paths cross again at a second beam splitter \textit{BS}$_2$.
At the exit of \textit{BS}$_2$ are two detectors, $D_X$ along the $x$
direction and $D_Y$ along the $y$ direction, as indicated in Fig.~\ref{MZ1}.

An explanation of the Mach-Zehnder interferometer is given in Ref.~\onlinecite{SS}.
For completeness and to establish the notation we summarize the
main physics of the Mach-Zehnder interferometer.
The state inside the interferometer can be modeled as a two-level system,
for instance, by associating the states $\s{x}$ and $\s{y}$ with the
particle at any time corresponding to its direction of propagation
inside the interferometer (see Fig.~\ref{MZ1}).

If both paths $A$ and $B$ have the same length, we can neglect the phase of the 
particle on the trajectories between the beam splitters and
mirrors. The effect of the interferometer on the
particle state is then given by the sequence of unitary transformations
$\mbox{\textit{BS}}_1 \to M_{A,B} \to \mbox{\textit{BS}}_2$.
A symmetric and equilibrated beam splitter is described by the unitary
transformation\cite{Z}
\begin{equation}
\mbox{Beam splitter:}
\begin{cases}
\s{x}\longrightarrow \dfrac{1}{\sqrt{2}} (\s{x}+i\s{y}),
\\
\s{y}\longrightarrow \dfrac{1}{\sqrt{2}} (\s{y}+i\s{x}),
\end{cases}
\end{equation}
while the combination of the mirrors $M_A$ and $M_B$ acts as $\s{x} \to i \s{y}$ and
$\s{y} \to i \s{x}$. We combine these transformation and see that the interferometer
acts as
\begin{equation} \label{eq:MZI_action}
\mbox{Mach-Zehnder interferometer:}
\begin{cases}
\s{x}\longrightarrow e^{i\pi}\s{x},
\\
\s{y}\longrightarrow e^{i\pi}\s{y}.
\end{cases}
\end{equation}
At the exit of the Mach-Zehnder interferometer the detector $D_X$ measures the
component $\s{x}$ of the outgoing state $\s{\psi_{\rm out}}$, and
$D_Y$, the component $\s{y}$ with the
probabilities
\begin{subequations}
\begin{align}
\p\{X\}&=\|P_{\s{x}}\s{\psi_{\rm out}}\|^2=|\langle x\s{\psi_{\rm out}}|^2, \\
\p\{Y\}&=\|P_{\s{y}}\s{\psi_{\rm out}}\|^2=|\langle y\s{\psi_{\rm out}}|^2,
\end{align}
\end{subequations}
where $P_{\s{x}}$ and $P_{\s{y}}$ are the projectors onto $\s{x}$ and $\s{y}$,
respectively.
Particles are injected from the source along the $x$ direction,
that is, in state $\s{x}$. From Eq. \eqref{eq:MZI_action} it therefore follows that
the probability of measuring the particle in detector $D_X$ is
$\p\{X\}=1$, and the probability of measuring it in detector $D_Y$
is $\p\{Y\}=0$. This result is in contrast to the expected classical result,
which is $\p\{X\}=\p\{Y\}=1/2$.
This effect is known as \emph{one-particle quantum interference}
and is one of the typical non-classical and counterintuitive
effects of the quantum physics.\cite{VS}
Increasing the length of one of the paths $A$ or $B$ leads to an additional
phase difference of the states before \textit{BS}$_2$ and can be used to control 
of the interference and hence the probabilities $\p\{X\}$
and $\p\{Y\}$.\cite{SS}

\section{Which-way detector} \label{WWMZdet}

\begin{figure}[h]
	\includegraphics[width=0.8\columnwidth]{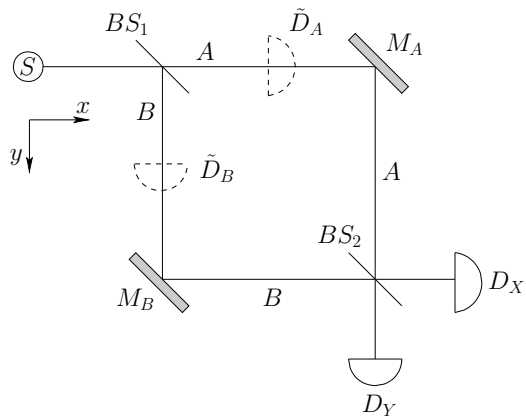}
\caption{Which-way detection in the Mach-Zehnder interferometer. Additional (nondestructive) detectors
$\tilde{D}_A$ and $\tilde{D}_B$ are placed on the paths $A$ and $B$.
The entanglement of the particle with $\tilde{D}_A$ and $\tilde{D}_B$ provides
information on which path was taken by it. This information destroys the
quantum interference and results in detection probabilities equal to 1/2 in $D_X$ and
$D_Y$ corresponding to the classical result.\label{WWMZ}}
\end{figure}

The quantum interference effect is destroyed if we put additional (nondestructive)
detectors $\tilde{D}_A$ on path $A$ and $\tilde{D}_B$ on path $B$
(see Fig.~\ref{WWMZ}) to detect which path was chosen by the
particle.
A detection by $\tilde{D}_A$ projects the particle onto the state $\s{x}$ and tells us that path $A$ was taken.
A detection by $\tilde{D}_B$ projects onto $\s{y}$ and tells us that path $B$ was taken.
Hence the
state at the exit of the interferometer is fully determined by the action of the beam splitter \textit{BS}$_2$ on
the incoming state from either path $A$ or path $B$.
The final beam splitting leads to the probabilities
$\p\{X\}=\frac12$ and $\p\{Y\}=\frac12$, equivalent to
the classical result and independent of the result detected by $\tilde{D}_X$ and $\tilde{D}_Y$. 
The quantum interference disappears, showing that
the concepts of ``quantum interference'' and ``knowledge of the path''
are complementary.

Note that the interference disappears as soon as the information on the path is 
stored in the system state. It is not a ``uncontrolled'' perturbation of the state 
of the quantum particle that destroys the interference.

\section{Which-way entangler}\label{WWMZent}

A remarkable property of quantum physics is that the interference vanishes
by the mere presence of the detectors $\tilde{D}_A$ and $\tilde{D}_B$, even
if the result of the measurement is not classically read out (which would correspond
to a projection on the path taken $A$ or $B$).
This property can be illustrated by a simple extension of the Mach-Zehnder interferometer, which also lets us
show that a detection takes place by \emph{entanglement} between the
particle and the detector. We call this model a \emph{which-way entangler}.
We start from the which-way detector shown in Fig.~\ref{WWMZ}.
In addition, we now assume that the particle is emitted into the
interferometer in an excited state $\s{e}$. The detection by detectors $\tilde{D}_A$ or
$\tilde{D}_B$ relaxes the particle into its ground state $\s{g}$ by emission of
a photon. We denote the photon state by $\s{A}$ or $\s{B}$, determined by which detector received the photon,
and use the notation $\s{0}$ for the absence of
any photon.

The state at the entrance of the interferometer is therefore
\begin{equation}
\s{\Psi_{\rm in}} = \s{x} \otimes \s{0} \otimes \s{e}.
\end{equation}
After the first beam splitter the state is
\begin{equation}
\s{\Psi_1} = \frac{1}{\sqrt{2}} \Big[\s{x}+i\s{y}\Big] \otimes \s{0} \otimes \s{e}.
\end{equation}
The action of the detectors $\tilde{D}_A$ or $\tilde{D}_B$ leads to the entangled state
\begin{equation}
\s{\Psi_2}
= \frac{1}{\sqrt{2}}
\Big[ \s{x} \otimes \s{A} + i \s{y} \otimes \s{B} \Big] \otimes \s{g}.
\label{above}
\end{equation}
Note that we do not classically read out the detectors here but keep the quantum coherent superposition
between the path $A$ and path $B$ detections by transferring the which-way information into the photon state.
As long as the photon state is not measured, the superposition is maintained.
This state becomes after the mirrors
\begin{equation}
\s{\Psi_3}
=
\frac{1}{\sqrt{2}} \Big[ i\s{y} \otimes \s{A} - \s{x} \otimes \s{B} \Big] \otimes \s{g},
\end{equation}
and as the final state after the second beam splitter
\begin{subequations}
\label{eq:psi_out_ent}
\begin{align}
\s{\Psi_{\rm out}}
&=
\frac{1}{2}
\Big[
\big(i\s{y}-\s{x}\big) \otimes \s{A}
-
\big(\s{x}+i\s{y}\bigr) \otimes \s{B}
\Big] \otimes \s{g}
\\
&=
\frac{1}{2}
\Big[
i\s{y} \otimes \big(\s{A}-\s{B}\big)
-
\s{x} \otimes \big(\s{A}+\s{B}\bigr)
\Bigr] \otimes \s{g}.
\end{align}
\end{subequations}
We see that even though we keep the superposition of the which-way results,
the interference effect at the final detectors $D_X$ and $D_Y$ is destroyed, and
the detection probabilities correspond to the classical results
\begin{subequations}
\begin{align}
\p\{X\}
&=
\|P_{\s{x}}\otimes I \otimes I \s{\Psi_{\rm out}}\|^2
=
\frac{1}{4}\|\s{A} + \s{B}\|^2
=
\frac{1}{2},
\\
\p\{Y\}
&=
\|P_{\s{y}} \otimes I \otimes I \s{\Psi_{\rm out}}\|^2
=
\frac{1}{4}\|\s{A} - \s{B}\|^2
=
\frac{1}{2}.
\end{align}
\end{subequations}
However, in addition to the which-way detection alone, we have
now transmitted the which-way information into the photon states
$\s{A}$ and $\s{B}$. Once the photon is emitted, it becomes \emph{entangled} with
the quantum particle state, which means that we can no longer write the state as a
simple product $\s{\Psi} = \s{\text{particle}} \otimes \s{\text{photon}}$,
as is clearly seen with $\s{\Psi_2}$ in Eq.~\eqref{above}.


\section{Quantum eraser}\label{WWMZera}

We now show that the which-way information can be erased in a simple way,
which restores the quantum interference at the output of the Mach-Zehnder interferometer.

\begin{figure}[h]
	\includegraphics[width=0.8\columnwidth]{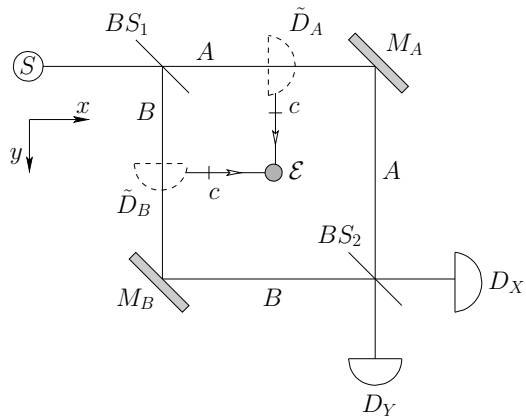}
\caption{Quantum eraser for the Mach-Zehnder interferometer. The which-way detection takes place
by emission of a photon in $\tilde{D}_A$ or $\tilde{D}_B$. If channel
$c$ is open, the photon can be absorbed, with some probability, by an auxiliary atom $\cal E$,
the quantum eraser. The absorption erases the entanglement with the detectors
and restores the quantum interference at the exit of the interferometer.\label{WWMZ1}}
\end{figure}

Consider the Mach-Zehnder interferometer, modified in the following way. We
assume that the photon after emission in one of the two $\tilde D$ detectors
travels to an auxiliary atom ${\cal E}$, the \emph{quantum eraser}, where it can be absorbed
(see Fig.~\ref{WWMZ1}).
To activate the quantum eraser, the observer has to open a channel $c$
connecting detectors $\tilde{D}_A$ and $\tilde{D}_B$ to atom ${\cal E}$.
This opening can be done at any time after the emission of the photons,
even when the quantum particle has left the Mach-Zehnder interferometer.\cite{ESW,M}

Let $\s{\gamma}$ be the initial (ground) state of the quantum eraser
$\cal E$ and let $\s{\varepsilon}$ be its excited state.
We consider the evolution of the system ``atom + photon + quantum eraser.''
If the channel $c$ is closed, then the state just before the final detection is
identical to Eq.~\eqref{eq:psi_out_ent} with the additional state $\s{\gamma}$
of ${\cal E}$,
\begin{subequations}
\begin{align}
	&\s{\Phi_{\rm out}}
	=
	\s{\Psi_{\rm out}}\otimes \s{\gamma}
	\\
	&=
	\frac{1}{2}
	\Big[
	i\s{y}\otimes \left(\s{A}-\s{B}\right)
	-
	\s{x} \otimes \left(\s{A}+\s{B}\right)
	\Big]
	\otimes
	\s{g} \otimes \s{\gamma},
\end{align}
\end{subequations}
and we obtain the same probabilities $1/2$ at the final detectors as in Sec.~IV.

In contrast, if channel $c$ is open, the photon travels from the detectors $\tilde{D}_A$ and $\tilde{D}_B$
to the eraser $\mathcal{E}$ where with some probability it can be absorbed by exciting the quantum eraser.
We stress that the absorption of the photon is probabilistic and depends on the
precise superposition of the $\s{A}$ and $\s{B}$ components of the photon state at $\mathcal{E}$
and on the cross-section of the absorption process. An absorption (erasure) that occurs with certainty
would be a nonunitary transformation which is forbidden by quantum physics
and would allow for paradoxes such as superluminous transmission of information, that is, a violation
of the no-signalling-theorem.\cite{GRW,B}

Therefore the quantum erasure occurs only for some outcomes of the interference experiments.
For those cases where the photon is absorbed by $\mathcal{E}$, the system state is projected onto
\begin{equation}\label{erasure}
\s{\Phi} = -\s{x} \otimes |0\rangle \otimes \s{g}\otimes \s{\varepsilon}.
\end{equation}
This state is identical (for the injected particle) to the usual action of the Mach-Zehnder interferometer
expressed by Eq.~\eqref{eq:MZI_action}, and therefore the single-particle interference
at the detectors $D_X$ and $D_Y$ is restored. However, the probabilities $\p\{X\}$
and $\p\{Y\}$ are now replaced by the conditional probabilities $\p\{X | \text{abs}\}$
and $\p\{Y | \text{abs}\}$, which involve the preselection of the measurements to only those
cases where the photon has actually been absorbed by the quantum eraser.\cite{GRW}
As a result we obtain
\begin{subequations}
\begin{align}
\p\{X | \text{abs}\}&= \|(P_{\s{x}}\otimes I\otimes I\otimes I)\s{\Phi}\|^2=1,\\
\p\{Y | \text{abs}\}&= \|(P_{\s{y}}\otimes I\otimes I\otimes I)\s{\Phi}\|^2=0.
\end{align}
\end{subequations}
We see that the erasure of the which-path information by the absorption of the photon by the quantum 
eraser completely restores the
original quantum interference.


\section{Conclusion}

We have presented a simple model that requires knowledge only
of two-level systems. Nonetheless, it allows us to explain interesting effects
about one-particle quantum interference:
Quantum interference appears when a particle can take different indistinguishable paths to
arrive at a detector.
The knowledge of which path was taken is obtained
by entanglement between the quantum particle and a detector on the path.
The loss of the one-particle quantum interference is an illustration
that this entanglement changes the state of the particle.
The interference can be restored by using the quantum eraser, which
disentangles the particle and detector states, and thus also erases any
which-way information.
Note that a noisy environment acts in a similar way as the which-path
detector and destroys the quantum interference by getting entangled
through interaction with
the particle. However, this effect is generally uncontrolled
and not reversible by a quantum eraser, and the result is a purely classically
operating Mach-Zehnder interferometer.

The complementarity between ``quantum interference'' and ``knowledge of the path''
in this simple model is manifestly evident:
``quantum interference'' corresponds to a factorized (product) state,
``knowledge of the path'' to an entangled state.


\begin{acknowledgments}
The authors thank V.\ Scarani for stimulating and helpful discussions, M.\
Bahrami for useful correspondence, and the anonymous referees of Am. J. Phys. for a
careful review and constructive comments.
B.B. acknowledges the support by the Swiss NCCR Nanoscience (Basel) and SNF.
\end{acknowledgments}



\end{document}